\documentclass{amsart}
\date{\today}
\usepackage{amssymb}
\usepackage{latexsym}
\usepackage{hyperref}

\newcommand{\Z}{{\mathbb Z}}
\newcommand{\R}{{\mathbb R}}
\newcommand{\C}{{\mathbb C}}

\newcommand{\T}{{\mathbb T}}
\newcommand{\E}{{\mathbb E}}
\newcommand{\PP}{{\mathbb P}}

\newtheorem{theorem}{Theorem}

\newtheorem{problem}{Problem}
\newtheorem{lemma}{Lemma}[section]
\newtheorem{prop}[lemma]{Proposition}
\newtheorem{coro}{Corollary}

\DeclareMathOperator{\supp}{supp}

\begin{document}

\title[A Survey of Kotani Theory]{Lyapunov Exponents and Spectral
Analysis of Ergodic Schr\"odinger Operators: A Survey of Kotani Theory and its
Applications}

\author{David Damanik}

\address{Mathematics 253--37, California Institute of
Technology, Pasadena, CA 91125, USA}

\email{\href{mailto:damanik@caltech.edu}{damanik@caltech.edu}}

\urladdr{\href{http://www.math.caltech.edu/people/damanik.html}{http://www.math.caltech.edu/people/damanik.html}}

\thanks{The author was supported in part by NSF grant DMS--0500910}



\begin{abstract}
The absolutely continuous spectrum of an ergodic family of one-dimensional Schr\"odinger
operators is completely determined by the Lyapunov exponent as shown by Ishii, Kotani and
Pastur.

Moreover, the part of the theory developed by Kotani gives powerful tools for proving the
absence of absolutely continuous spectrum, the presence of absolutely continuous
spectrum, and even the presence of purely absolutely continuous spectrum.

We review these results and their recent applications to a number of problems: the
absence of absolutely continuous spectrum for rough potentials, the absence of absolutely
continuous spectrum for potentials defined by the doubling map on the circle, and the
absence of singular spectrum for the subcritical almost Mathieu operator.
\end{abstract}

\dedicatory{Dedicated to Barry Simon on the occasion of his 60th birthday.}

\maketitle

%
%
%
%

\section{Introduction}

Schr\"odinger operators with ergodic potentials have enjoyed quite some popularity for
several decades now. This is in no small part due to Barry Simon's contributions to the
field, through research articles on the one hand, but also through survey articles and
his way of putting his personal stamp on results and conjectures on the other hand.
Ergodic Schr\"odinger operators continue to be dear to him as seven of the fifteen
Schr\"odinger operator problems he singles out in \cite{s21st} for further investigation
in the 21st century deal with them. Moreover, the immense activity in the area of ergodic
Schr\"odinger operators is reflected by the fact that the ratio $7/15$ improves to $3/4$
when it comes to the problems from that list that have been solved so far. One may say
that this is due to the uneven distribution of difficulty among these fifteen problems,
but this is balanced by the fact that of the remaining four ergodic problems at least
three are very hard and that further progress should be expected on some of the remaining
non-ergodic problems.

In the area of ergodic Schr\"odinger operators there are several powerful methods (e.g.,
fractional moment) and analyses (multi-scale) but few theories (Kotani). What appears to
be wordplay wants to express the fact that Kotani theory is distinguished from the other
greats by its immensely general scope. It really is a theory that applies to the class of
all ergodic operators and it is central in many ways. In addition, Kotani theory has
played a crucial role in the solution of two of the four recently solved 21st century
problems.

Our goal here is to present the core parts of Kotani theory with more or less complete
proofs and to discuss several recent applications of the theory to a number of concrete
classes of models for which, whenever possible, we at least outline the main ideas that
go into the proofs of the results we mention.

Suppose $(\Omega,\mu)$ is a probability measure space, $T : \Omega \to \Omega$ is an
invertible ergodic transformation, and $f : \Omega \to \R$ is bounded and measurable. We
define potentials,
$$
V_\omega(n) = f(T^n \omega), \quad \omega \in \Omega, \quad n \in \Z,
$$
and the corresponding discrete Schr\"odinger operators in $\ell^2(\Z)$,
\begin{equation}\label{oper}
[H_\omega \psi](n) = \psi(n+1) + \psi(n-1) + V_\omega(n) \psi(n).
\end{equation}
We will call $\{H_\omega\}_{\omega \in \Omega}$ an ergodic family of Schr\"odinger
operators.
\\[3mm]
\textit{Examples.} (a) Quasi-periodic potentials: $\Omega = \T^d = \R^d / \Z^d$, $\mu$ is
the normalized Lebesgue measure on $\T^d$, and $T \omega = \omega + \alpha$ is some
ergodic shift (i.e., $1,\alpha_1,\ldots,\alpha_d$ are rationally
independent).\footnote{What we call quasi-periodic here is more general than the notion
of quasi-periodicity as defined in \cite{as3}, for example, where a quasi-periodic
potential is almost periodic with a finitely generated frequency module. In particular, a
quasi-periodic potential as defined here is not necessarily almost periodic, that is, the
translates of a given quasi-periodic potential are not necessarily precompact in $\ell^\infty(\Z)$.
We allow discontinuous sampling functions $f$ here because we want to include Fibonacci-type
potentials.}\\
(b) Potentials defined by the skew shift: $\Omega = \T^2$, $\mu$ is the normalized
Lebesgue measure on $\T^2$, and $T (\omega_1,\omega_2) = (\omega_1 + \omega_2, \omega_2 +
\alpha)$
for some irrational $\alpha$.\\
(c) Potentials defined by the doubling map: $\Omega = \T$, $\mu$ is the normalized
Lebesgue measure on $\T$, and $T \omega = 2 \omega$.\footnote{Strictly speaking, this
example does not fall within our general framework as $T$ is not invertible, but
potentials of this kind have been studied in several works and it is possible to tweak
the model a little to
fit it in the framework above.}\\
(d) Potentials defined by the left shift: $\Omega = I^\Z$, where $I$ is a compact subset
of $\R$, $\mu = \PP^\Z$, where $\PP$ is a Borel probability measure on $I$, and $[T
\omega](n) = \omega(n+1)$.
\\[3mm]
The following pair of results, proven in \cite{ks2,p}, shows that for ergodic families of
Schr\"odinger operators, the spectrum and the spectral type are deterministic in the
sense that they are constant $\mu$-almost surely.

\begin{theorem}[Pastur 1980]
There exists a set $\Sigma \subset \R$ such that for $\mu$-almost every $\omega$,
$\sigma(H_\omega) = \Sigma$.
\end{theorem}

\begin{theorem}[Kunz-Souillard 1980]
There are sets $\Sigma_\mathrm{ac}$, $\Sigma_\mathrm{sc}$, $\Sigma_\mathrm{pp} \subset
\R$ such that for $\mu$-almost every $\omega$, $\sigma_\bullet(H_\omega) =
\Sigma_\bullet$, $\bullet \in \{ \mathrm{ac,sc,pp} \}$.
\end{theorem}

Thus, in the spectral analysis of a given ergodic family of Schr\"odinger operators, a
fundamental problem is the identification of the sets $\Sigma$, $\Sigma_\mathrm{ac}$,
$\Sigma_\mathrm{sc}$, and $\Sigma_\mathrm{pp}$.

The almost sure spectrum, $\Sigma$, is completely described by the integrated density of
states as shown by Avron and Simon \cite{as}. Denote the restriction of $H_\omega$ to
$[0,N-1]$ with Dirichlet boundary conditions by $H_{\omega}^{(N)}$. For $\omega \in
\Omega$ and $N \ge 1$, define measures $dk_{\omega,N}$ by placing uniformly distributed
point masses at the eigenvalues $E^{(N)}_\omega (1) < \cdots < E^{(N)}_\omega (N)$ of
$H_\omega^{(N)}$, that is,
$$
\int f(E) \, dk_{\omega,N}(E) = \frac{1}{N} \, \sum_{n = 1}^N f(E^{(N)}_\omega (n)).
$$
Then, it can be shown that for $\mu$-almost every $\omega \in \Omega$, the measures
$d\tilde{k}_{\omega,N}$ converge weakly to a non-random measure $dk$, called the
\textit{density of states measure}, as $N \to \infty$. The function $k$ defined by
$$
k(E) = \int \chi_{(-\infty,E]} (E') \, dk(E')
$$
is called the \textit{integrated density of states}. It is not hard to show that
\begin{equation}\label{idsaverage}
\int f(E) \, dk(E) = \E \left( \langle \delta_0 , f(H_\omega) \delta_0 \rangle \right)
\end{equation}
for bounded measurable $f$. Here, $\E ( \cdot )$ denotes integration with respect to the
measure $\mu$, that is, $\E(g) = \int g(\omega) \, d\mu(\omega)$ Thus, the density of
states measure is given by an average of the spectral measures associated with $H_\omega$
and $\delta_0$. The $T$-invariance of $\mu$ then implies the following result:

\begin{theorem}[Avron-Simon 1983]\label{avsimthm}
The almost sure spectrum is given by the points of increase of $k$, that is, $\Sigma =
\mathrm{supp} (dk)$.
\end{theorem}

There is a similarly general description of the set $\Sigma_\mathrm{ac}$ in terms of the
Lyapunov exponent. Let $E \in \C$ and
\begin{equation}\label{schcoc}
A^E(\omega) = \left( \begin{array}{cr} E - f(\omega) & -1 \\ 1 & 0 \end{array} \right) .
\end{equation}
Define the \textit{Lyapunov exponent} $\gamma(E)$ by
$$
\gamma (E) = \lim_{n \to \infty} \frac{1}{n} \E \left( \log \| A^E_n (\omega) \| \right),
$$
where $A^E_n (\omega) = A^E (T^{n-1} \omega) \cdots A^E (\omega)$.

The integrated density of states and the Lyapunov exponent are related by the Thouless
formula (see, e.g., \cite[Theorem~9.20]{cfks}), which reads
\begin{equation}\label{thouless}
\gamma(E) = \int \log | E - E' | \, dk(E').
\end{equation}

The significance of the transfer matrices $A^E_n(\omega)$ is that a sequence $u$ solves
the difference equation
\begin{equation}\label{eve}
u(n+1) + u(n-1) + V_\omega(n) u(n) = E u(n)
\end{equation}
if and only if
$$
\left( \begin{matrix} u_n \\ u_{n-1}
\end{matrix} \right) = A^E_n (\omega) \left( \begin{matrix} u_0 \\ u_{-1}
\end{matrix} \right)
$$
for every $n$. Since $\det A^E(\omega) = 1$, we always have $\gamma(E) \ge 0$. Let us
define
$$
\mathcal{Z} = \{ E : \gamma(E) = 0 \}
$$
By general principles, $\mathcal{Z} \subseteq \Sigma$.

For a subset $S$ of $\R$, the essential closure of $S$ is given by
$$
\overline{S}^\mathrm{ess} = \{ E \in \R : \mathrm{Leb}(S \cap
(E-\varepsilon,E+\varepsilon)) > 0 \text{ for every } \varepsilon > 0 \}
$$
Then, the following theorem combines results from \cite{i,k1,p}.\footnote{To be more
precise, the discrete version of the Kotani half of this result here can be found in the
paper \cite{s} by Simon and the work of Ishii and Pastur was preceded by closely related
work by Casher and Lebowitz \cite{cashl}.}

\begin{theorem}[Ishii 1973, Pastur 1980, Kotani 1984]\label{ipkthm}
The almost sure absolutely continuous spectrum is given by the essential closure of the
set of energies for which the Lyapunov exponent vanishes, that is, $\Sigma_\mathrm{ac} =
\overline{\mathcal{Z}}^\mathrm{ess}$.
\end{theorem}

While there is an analog of Theorem~\ref{avsimthm} for higher-dimensional ergodic
Schr\"odinger operators, Theorem~\ref{ipkthm} is, by its very nature, a strictly
one-dimensional result. It is one of the great challenges for researchers in the area of
ergodic Schr\"odinger operators to develop effective tools for the study of the
absolutely continuous spectrum in higher dimensions. That said, Theorem~\ref{ipkthm}
holds in virtually all one-dimensional and quasi-one-dimensional situations: see Kotani
\cite{k1} for continuous one-dimensional Schr\"odinger operators (see also Kirsch
\cite{k0} for a useful extension), Minami \cite{m} for generalized Sturm-Liouville
operators, Kotani-Simon \cite{ks} for discrete and continuous Schr\"odinger operators
with matrix-valued potentials, and Geronimo \cite{g0} and Geronimo-Teplyaev \cite{gt} for
orthogonal polynomials on the unit circle.

%
%
%
%

\section{The Description of the AC Spectrum}

In this section we discuss the main ideas that go into the proof of Theorem~\ref{ipkthm}.
The proof naturally breaks up into the proof of two inclusions.

The inclusion ``$\subseteq$'' was proved by Ishii and Pastur. The other inclusion was
proved by Kotani and is a much deeper result. In fact, the proof of the Ishii-Pastur half
of the result we give below is based on more modern techniques and shows that this half
is really an immediate consequence of the general theory of one-dimensional Schr\"odinger
operators.

There are at least three different proofs of the Ishii-Pastur half of
Theorem~\ref{ipkthm} in the literature. One of them uses the existence of generalized
eigenfunctions; compare Cycon et al.\ \cite{cfks}. The second one, due to Deift and Simon
\cite{ds}, is close in spirit to, and uses techniques from, Kotani's proof of the other
half of the result. Finally, there are two somewhat related proofs given by Buschmann
\cite{busch} and Last and Simon \cite{ls}, which are both either directly or indirectly
based on a result of Gilbert and Pearson that describes a support of the singular
spectrum of a Schr\"odinger operator with fixed (non-random) potential in terms of
subordinate solutions. We will follow the argument from Buschmann's paper below.

We first recall a central result from Gilbert and Pearson's subordinacy theory
\cite{g,gp}. Consider the discrete Schr\"odinger operator $H$ in $\ell^2(\Z)$ with
potential $V$ and solutions of the difference equation
\begin{equation}\label{evedet}
u(n+1) + u(n-1) + V(n)u(n) = E u(n).
\end{equation}
A non-zero solution $u$ of \eqref{evedet} is called \textit{subordinate} at $\pm \infty$
if for every linearly independent solution $\tilde u$ of \eqref{evedet}, we have
$$
\frac{\sum_{n = 0}^{N-1} |u(\pm n)|^2}{\sum_{n = 0}^{N-1} |\tilde u(\pm n)|^2} \to 0
\text{ as } N \to \infty.
$$
Let
$$
S = \{ E \in \R : \text{ \eqref{evedet} has solutions $u_+$ and $u_-$ such that $u_\pm$
is subordinate at } \pm \infty \}.
$$
Then, $S$ has zero weight with respect to the absolutely continuous part of any spectral
measure, that is,
\begin{equation}\label{gilbert}
\mathcal{P}^{\mathrm{(ac)}} (S) = 0.
\end{equation}

\begin{proof}[Proof of the Ishii-Pastur half of Theorem~\ref{ipkthm}.] Note that when
$\gamma(E) > 0$, Oseledec' Theorem \cite{o} says that for almost every
$\omega$, there are solutions $u_+(E,\omega)$ and $u_-(E,\omega)$ of \eqref{eve} such
that $u_\pm (E,\omega)$ is exponentially decaying, and hence subordinate, at $\pm
\infty$. Applying Fubini's theorem, we see that for $\mu$-almost every $\omega$, the set
of $E \in \R \setminus \mathcal{Z}$ for which the property just described fails, has zero
Lebesgue measure. In other words, for these $\omega$'s, $\R \setminus \mathcal{Z}
\subseteq S_\omega$ up to a set of zero Lebesgue measure. Since sets of zero Lebesgue
measure have zero weight with respect to the absolutely continuous part of any spectral
measure, we obtain from \eqref{gilbert} that for $\mu$-almost every $\omega$,
$$
\mathcal{P}_\omega^{\mathrm{(ac)}} (\R \setminus \mathcal{Z}) = 0.
$$
This shows that for $\mu$-almost every $\omega$, $\sigma_\mathrm{ac}(H_\omega) \subseteq
\overline{\mathcal{Z}}^\mathrm{ess}$.
\end{proof}

Let us now turn to the Kotani half of Theorem~\ref{ipkthm}. Kotani worked in the
continuum setting. Carrying his results over to the discrete case is not entirely
straightforward and it was worked out by Simon \cite{s} whose proof we give below. Given
$z \in \C_+ = \{ z \in \C : \Im z > 0 \}$ and $\omega \in \Omega$, there are (up to a
multiplicative constant) unique solutions $u_\pm(n,\omega)$ of
\begin{equation}\label{evez}
u(n+1) + u(n-1) + V_\omega(n) u(n) = z u(n)
\end{equation}
such that $u_\pm$ is square-summable at $\pm \infty$. (Take $u_\pm(n,\omega) = \langle
\delta_n, (H_\omega - z)^{-1} \delta_1 \rangle$ near $\pm \infty$ to show existence;
uniqueness follows from constancy of the Wronskian.) Note that $u_\pm(0,\omega) \not= 0$
for otherwise $z$ would be a non-real eigenvalue of a self-adjoint half-line operator.
Thus, we can define
\begin{equation}\label{mpmdef}
m_\pm(z,\omega) = - \frac{u_\pm(\pm 1,\omega)}{u_\pm(0,\omega)}.
\end{equation}
Clearly,
\begin{equation}\label{shiftedm}
m_\pm(z,T^n \omega) = - \frac{u_\pm(n \pm 1,\omega)}{u_\pm(n,\omega)}.
\end{equation}

By Oseledec' Theorem, we have for $\mu$-almost every $\omega$,
$$
\lim_{n \to \infty} \frac1n \log \left| \frac{u_\pm(n,\omega)}{u_\pm(0,\omega)} \right| =
- \gamma(z).
$$
By \eqref{shiftedm},
$$
\log \left| \frac{u_\pm(n,\omega)}{u_\pm(0,\omega)} \right| = \sum_{m = 0}^{n-1} \log |
m_\pm(z,T^{\pm m} \omega) |
$$
and hence Birkhoff's ergodic theorem implies
\begin{equation}\label{emgamma}
\E (\log |m_\pm (z,\omega)|) = - \gamma(z).
\end{equation}

\begin{prop}\label{elogmgamma}
We have that
$$
\E \left( \log \left( 1 + \frac{\Im z}{\Im m_\pm (z,\omega)} \right) \right) = 2
\gamma(z).
$$
\end{prop}

\begin{proof}
By the difference equation \eqref{evez} that $u_\pm$ obeys,
\begin{equation}\label{riccati}
m_\pm(z,T^{n}\omega) = V_\omega(n) - z - [m_\pm(z,T^{n \mp 1} \omega)]^{-1}.
\end{equation}
Taking imaginary parts,
$$
\Im m_\pm(z,\omega) = - \Im z - \Im \left( [m_\pm(z,T^{\mp 1}\omega)]^{-1} \right).
$$
Dividing by $\Im m_+(z,\omega)$,
$$
1 =  - \frac{\Im z}{\Im m_\pm(z,\omega)} - \frac{\Im \left( [m_\pm(z,T^{\mp
1}\omega)]^{-1} \right)}{\Im m_\pm(z,\omega)}.
$$
Taking the logarithm,
\begin{align*}
\log \left( 1 + \frac{\Im z}{\Im m_\pm(z,\omega)} \right) & = \log \left( - \Im
\left([m_\pm(z,T^{\mp 1}\omega)]^{-1} \right) \right) - \log \left( \Im m_\pm(z,\omega) \right) \\
& = \log \left( \frac{\Im m_\pm(z,T^{\mp 1}\omega)}{|m_\pm(z,T^{\mp 1}\omega)|^2} \right)
- \log \left( \Im m_\pm(z,\omega) \right)
\end{align*}
Taking expectations and using invariance,
\begin{align*}
\E \left( \log \left( 1 + \frac{\Im z}{\Im m_\pm(z,\omega)} \right) \right) & = \E \left(
\log \left( \frac{\Im m_\pm(z,T^{\mp 1}\omega)}{|m_\pm(z,T^{\mp 1}\omega)|^2}
\right) - \log \left( \Im m_\pm(z,\omega) \right) \right) \\
& = - 2 \, \E \left( \log |m_\pm(z,\omega)| \right) \\
& = 2 \gamma(z),
\end{align*}
where we used \eqref{emgamma} in the last step.
\end{proof}

Denote
$$
b(z,\omega) = m_+(z,\omega) + m_-(z,\omega) + z - V_\omega(0).
$$

\begin{prop}\label{prop63}
We have
$$
\E \left( \Im \left( \frac{1}{b(z,\omega)} \right) \right) = - \frac{\partial
\gamma(z)}{\partial (\Im z)} .
$$
\end{prop}

\begin{proof}
It follows from \eqref{riccati} that
\begin{equation}\label{mminus}
\frac{u_-(1,\omega)}{u_-(0,\omega)} = m_-(z,\omega) + z - V_\omega(0).
\end{equation}
It is not hard to check that for $n \le m$,
\begin{equation}\label{green}
G_\omega(n,m;z) : = \langle \delta_n, (H_\omega - z)^{-1} \delta_m \rangle =
\frac{u_-(n,\omega) u_+(m,\omega)}{u_+(1,\omega) u_-(0,\omega) - u_-(1,\omega)
u_+(0,\omega)}.
\end{equation}
From \eqref{mpmdef}, \eqref{mminus}, \eqref{green}, we get
\begin{equation}\label{greenmm}
-G_\omega(0,0;z)^{-1} = m_+(z,\omega) + m_-(z,\omega) + z - V_\omega(0) = b(z,\omega).
\end{equation}
The definition of $G_\omega(n,m;z)$ gives
\begin{equation}\label{greenids}
\E ( G_\omega(0,0;z) ) = \int \frac{1}{E' - z} \, dk(E').
\end{equation}
Thus,
\begin{align*}
\E \left( \Im \left( \frac{1}{b(z,\omega)} \right) \right) & = - \Im \E \left( G_\omega(0,0;z) \right) \\
& = - \Im \int \frac{1}{E' - z} \, dk(E') \\
& = - \frac{\partial}{\partial (\Im z)} \int \log |z - E'| \, dk(E')\\
& = - \frac{\partial \gamma(z)}{\partial (\Im z)},
\end{align*}
where we used \eqref{greenmm}, \eqref{greenids}, and the Thouless formula
\eqref{thouless}.
\end{proof}

Denote
$$
n_\pm(z,\omega) = \Im m_\pm (z,\omega) + \tfrac12 \Im z.
$$

\begin{prop}
We have that
\begin{equation}\label{propparta}
\E \left( \frac{1}{n_\pm(z,\omega)} \right) \le \frac{2 \gamma(z)}{\Im z}
\end{equation}
and
\begin{equation}\label{proppartb}
\E \left( \frac{ \left[ \frac{1}{n_+} + \frac{1}{n_-} \right] \cdot \left[ (n_+ - n_-)^2
+ (\Re b)^2 \right] }{|b|^2} \right) \le 4 \left[ \frac{\gamma(z)}{\Im z} -
\frac{\partial \gamma(z)}{\partial (\Im z)} \right].
\end{equation}
\end{prop}

\begin{proof}
For $x \ge 0$, consider the function $A(x) = \log (1 + x) - \frac{x}{1 + \frac{x}{2}}$.
Clearly, $A(0) = 0$ and $A'(x) = \frac{1}{1 + x} - \frac{1}{1 + x + \frac{x^2}{4}} \ge
0$. Therefore,
\begin{equation}\label{littleobs}
\log (1 + x) \ge \frac{x}{1 + \frac{x}{2}} \quad \text{ for all } x \ge 0.
\end{equation}
Thus,
\begin{align*}
\E \left( \frac{1}{n_\pm(z,\omega)} \right) & = \E \left( \frac{1}{\Im m_\pm(z,\omega) +
\tfrac12 \Im z} \right) \\
& = \frac{1}{\Im z} \, \E \left( \frac{\frac{\Im z}{\Im m_\pm(z,\omega)}}{1 +
\frac{\frac{\Im z}{\Im m_\pm(z,\omega)}}{2}} \right) \\
& \le \frac{1}{\Im z} \, \E \left( \log \left( 1 + \frac{\Im z}{\Im m_\pm(z,\omega)} \right) \right) \\
& = \frac{2 \gamma(z)}{\Im z},
\end{align*}
which is \eqref{propparta}. We used \eqref{littleobs} in the third step and
Proposition~\ref{elogmgamma} in the last step.

Notice that $n_+(z,\omega) + n_-(z,\omega) = \Im b(z,\omega)$. Thus the integrand on the
left-hand side of \eqref{proppartb} is equal to
\begin{align*}
\frac{ \left[ \cfrac{1}{n_+} + \cfrac{1}{n_-} \right] \cdot \left[ (n_+ + n_-)^2 - 4 n_+
n_- + (\Re b)^2 \right] }{|b|^2} & = \frac{ \left[ \cfrac{1}{n_+} + \cfrac{1}{n_-}
\right]
\cdot \left[ |b|^2 - 4 n_+ n_- \right] }{|b|^2} \\
& = \frac{1}{n_+} + \frac{1}{n_-} - 4 \, \frac{n_+ + n_-}{|b|^2} \\
& = \frac{1}{n_+} + \frac{1}{n_-} + 4 \, \Im \left( \frac{1}{b} \right).
\end{align*}
The bound \eqref{proppartb} now follows from \eqref{propparta} and
Proposition~\ref{prop63}.
\end{proof}

\begin{proof}[Proof of the Kotani half of Theorem~\ref{ipkthm}.]
The Thouless formula \eqref{thouless} says that
$$
\gamma(z) = \int \log | z - E' | \, dk(E') = \Re \int \log ( z - E' ) \, dk(E')
$$
and hence $- \gamma(z)$ is the real part of a function whose derivative is a Borel
transform (namely, of the measure $dk$). By general properties of the Borel transform, it
follows that the limit $\gamma'(E+i0)$ exists for Lebesgue almost every $E \in \R$ and,
in particular, for almost every $E \in \mathcal{Z}$. For these $E$, we have that
\begin{equation}\label{ggprime}
\lim_{\varepsilon \downarrow 0} \frac{\gamma (E + i \varepsilon)}{\varepsilon} =
\lim_{\varepsilon \downarrow 0} \frac{\gamma (E + i \varepsilon) - \gamma(E)}{\varepsilon
- 0} = \lim_{\varepsilon \downarrow 0} \frac{\partial \gamma}{\partial (\Im z)} (E + i
\varepsilon),
\end{equation}
and in particular, the limit is finite. Thus, by \eqref{propparta},
$$
\limsup_{\varepsilon \downarrow 0} \E \left( \frac{1}{\Im m_\pm(E+i\varepsilon,\omega)}
\right) < \infty
$$
for almost every $E \in \mathcal{Z}$. Since $m_\pm$ are Borel transforms as well (of the
spectral measures associated with half-line restrictions of $H_\omega$), we also have
that, for every $\omega \in \Omega$, $m_\pm(E+i0, \omega)$ exists for Lebesgue almost
every $E \in \R$, and hence, for almost every $E$, $m_\pm(E+i0, \omega)$ exists for
almost every $\omega$. Combining the last two observations with Fatou's lemma, we find
that
\begin{equation}\label{oneimbound}
\E \left( \frac{1}{\Im m_\pm(E+i0,\omega)} \right) < \infty
\end{equation}
for almost every $E$ in $\mathcal{Z}$. So, for almost every $\omega \in \Omega$ and $E
\in \mathcal{Z}$, $\Im m_\pm(E+i0,\omega) > 0$.

On the other hand, $m_+(E + i \varepsilon,\omega) + m_-(E + i \varepsilon,\omega) + E + i
\varepsilon - V_\omega(0)$ has a finite limit for almost every $\omega \in \Omega$ and $E
\in \mathcal{Z}$.

Hence, \eqref{greenmm} shows that $0 < \Im G_\omega(0,0;E+i0) < \infty$ for almost every
$\omega \in \Omega$ and $E \in \mathcal{Z}$, which implies the result.
\end{proof}

Denote the measure associated with the Herglotz function $G_\omega(0,0;z)$ by
$\nu_\omega$, that is,
$$
G_\omega(0,0;z) = \int \frac{d \nu_\omega(E)}{E - z}.
$$
The results above imply the following for $\mu$-almost every $\omega$:
\begin{align*}
\nu_\omega^{(\mathrm{ac})}(E) = 0 & \text{ for Lebesgue almost every } E \in \R \setminus
\mathcal{Z},\\
\nu_\omega^{(\mathrm{ac})}(E) > 0 & \text{ for Lebesgue almost every } E \in \mathcal{Z}.
\end{align*}
Here, $\nu_\omega^{(\mathrm{ac})}(E)$ denotes the density of the absolutely continuous
part of $\nu_\omega$. Write $k^{(\mathrm{ac})}(E)$ for the density of the absolutely
continuous part of the density of states measure.

There is a direct relation between these densities \cite{k5}:

\begin{theorem}[Kotani 1997]
For almost every $E \in \mathcal{Z}$,
\begin{equation}\label{acpartids}
k^{(\mathrm{ac})}(E) = \E \left( \nu_\omega^{(\mathrm{ac})}(E) \right).
\end{equation}
\end{theorem}

\begin{proof}
The inequality ``$\ge$'' in \eqref{acpartids} follows from \eqref{idsaverage} (i.e., the
density of states measure is the average of the measures $\nu_\omega$) and the fact that
the average of absolutely continuous measures is absolutely continuous.

To prove the opposite inequality, we first note that for almost every $E \in
\mathcal{Z}$, \eqref{greenids}, \eqref{ggprime}, and Cauchy-Riemann imply
\begin{equation}\label{lekac}
k^{(\mathrm{ac})}(E) = \frac1\pi \lim_{\varepsilon \downarrow 0} \frac{\gamma(E + i
\varepsilon)}{\varepsilon}.
\end{equation}

Because of \eqref{ggprime}, \eqref{oneimbound} and Fatou's lemma, \eqref{proppartb}
implies that for almost every pair $(E,\omega) \in \mathcal{Z} \times \Omega$,
\begin{equation}\label{immpmeq}
\Im m_+ (E+i0,\omega) = \Im m_- (E+i0, \omega)
\end{equation}
and
\begin{equation}\label{rempmeq}
\Re m_+ (E+i0,\omega) + \Re m_- (E+i0, \omega) + E - V_\omega(0) = 0.
\end{equation}
Thus, for almost every $(E,\omega) \in \mathcal{Z} \times \Omega$,
\begin{align}
\nu_\omega^{(\mathrm{ac})}(E) & = \frac{1}{\pi} \Im G_\omega(0,0;E + i0) \label{hm}\\
& = \frac{1}{\pi} \Im \frac{-1}{m_+(E + i0,\omega) + m_-(E
+ i0,\omega) + E - V_\omega(0)} \nonumber \\
& = \frac{1}{\pi} \Im \frac{-1}{2 i \Im m_+ (E+i0,\omega)} \nonumber \\
& = \frac{1}{2\pi} \frac{1}{\Im m_+ (E+i0,\omega)} \nonumber
\end{align}

Let $P_\varepsilon$ be the Poisson kernel for the upper half-plane, that is,
$$
P_\varepsilon(E) = \frac{1}{\pi} \, \frac{\varepsilon}{E^2 + \varepsilon^2}.
$$
Write
$$
C_\varepsilon(E) = \int_\mathcal{Z} P_\varepsilon(E - E') \, dE'
$$
and
$$
\tilde P_\varepsilon(E,E') = P_\varepsilon(E - E') C_\varepsilon(E)^{-1}.
$$
Then, by \eqref{hm} and Jensen's inequality, we obtain for almost every $(E,\omega) \in
\mathcal{Z} \times \Omega$,
\begin{align*}
\int_\R \nu_\omega^{(\mathrm{ac})}(E') & \, P_\varepsilon(E - E') \, dE' \ge
\int_\mathcal{Z} \nu_\omega^{(\mathrm{ac})}(E') \, P_\varepsilon(E
- E') \, dE' \\
& = \int_\mathcal{Z} \left( \frac{1}{2\pi} \frac{1}{\Im m_+ (E' + i0,\omega)} \right) \,
P_\varepsilon(E - E') \, dE' \\
& = C_\varepsilon(E) \int_\mathcal{Z} \left( \frac{1}{2\pi} \frac{1}{\Im m_+
(E' + i0,\omega)} \right) \, \tilde P_\varepsilon(E,E') \, dE' \\
& \ge C_\varepsilon(E) \left( \int_\mathcal{Z}\left( \frac{1}{2\pi} \frac{1}{\Im m_+
(E' + i0,\omega)} \right)^{-1} \tilde P_\varepsilon(E,E') \, dE' \right)^{-1} \\
& \ge C_\varepsilon(E)^2 \left( \int_\R \left( \frac{1}{2\pi} \frac{1}{\Im m_+
(E' + i0,\omega)} \right)^{-1} P_\varepsilon(E - E') \, dE' \right)^{-1} \\
& = \frac{C_\varepsilon(E)^2}{2\pi} \frac{1}{\Im m_+ (E+i\varepsilon,\omega)}
\end{align*}

Thus, for almost every $E \in \mathcal{Z}$,
$$
\int_\R \E \left( \nu_\omega^{(\mathrm{ac})}(E') \right)  P_\varepsilon(E - E') \, dE'
\ge \frac{C_\varepsilon(E)^2}{2\pi} \, \E \left( \frac{1}{\Im m_+
(E+i\varepsilon,\omega)} \right) ,
$$
and hence
\begin{equation}\label{hm2}
\E \left( \nu_\omega^{(\mathrm{ac})}(E) \right) \ge \frac{1}{2\pi} \limsup_{\varepsilon
\downarrow 0} \E \left( \frac{1}{\Im m_+ (E+i\varepsilon,\omega)} \right)
\end{equation}
since $C_\varepsilon(E) < 1$ and $C_\varepsilon(E) \to 1$ as $\varepsilon \downarrow 0$.

Using \eqref{lekac}, Proposition~\ref{elogmgamma}, the inequality $\log (1+x) \le x$ for
$x \ge 0$, and then \eqref{hm2}, we find that
\begin{align*}
k^{(\mathrm{ac})}(E) & = \frac1\pi \lim_{\varepsilon \downarrow 0} \frac{\gamma(E + i
\varepsilon)}{\varepsilon} \\
& = \frac1\pi \lim_{\varepsilon \downarrow 0} \frac{1}{2\varepsilon} \, \E \left( \log
\left( 1 + \frac{\varepsilon}{\Im m_\pm (E + i \varepsilon,\omega)} \right) \right) \\
& \le \frac{1}{2\pi} \limsup_{\varepsilon \downarrow 0} \E \left( \frac{1}{\Im m_\pm (E +
i \varepsilon,\omega)} \right) \\
& \le \E \left( \nu_\omega^{(\mathrm{ac})}(E) \right),
\end{align*}
concluding the proof of ``$\le$'' in \eqref{acpartids}.
\end{proof}

\begin{coro}\label{pureaccoro}
The spectrum is almost surely purely absolutely continuous if and only if the integrated
density of states is absolutely continuous and the Lyapunov exponent vanishes almost
everywhere with respect to the density of states measure.
\end{coro}

There is a different approach to purely absolutely continuous spectrum as pointed out by
Yoram Last (unpublished):\footnote{The author is grateful to Barry Simon for bringing
this to his attention.} Using a result of Deift and Simon \cite[Theorem~7.1]{ds}, one can
show that there is a set $\mathcal{R} \subseteq \mathcal{Z}$ such that
$\mathrm{Leb}(\mathcal{Z} \setminus \mathcal{R}) = 0$ and $\mathcal{R}$ has zero singular
spectral measure for every $\omega \in \Omega$ due to the absence of subordinate
solutions.

The spectrum naturally breaks up into the two components $\mathcal{Z}$ and $\Sigma
\setminus \mathcal{Z}$. It is known that either set can support singular continuous
spectrum as demonstrated by the almost Mathieu operator at critical and super-critical
coupling. However, as we have seen, $\Sigma \setminus \mathcal{Z}$ does not support any
absolutely continuous part of the spectral measures. Trivially, Anderson localization is
impossible in $\mathcal{Z}$. However, it is tempting to expect even more:

\begin{problem}
Prove or disprove that for all ergodic families, the operators $H_\omega$ have no
eigenvalues in $\mathcal{Z}$.
\end{problem}

If the answer is affirmative, this will in particular imply the absence of eigenvalues in
a number of special cases, such as the operators considered in Section~\ref{secfv} and
the critical almost Mathieu operator.\footnote{For this particular model, this would
provide an alternative to the proof based on self-duality and zero-measure spectrum.}

%
%
%
%

\section{The Induced Measure and its Topological Support}

Fix a compact subset $R$ of $\R$. Endow $R^\Z$ with product topology, which makes it a
compact metric space. If $V \in R^\Z$, we define the functions $m_\pm$ by
$$
m_\pm(z) = \mp \frac{u_\pm(1)}{u_\pm(0)},
$$
where $u_\pm$ solves
\begin{equation}\label{evedetz}
u(n+1) + u(n-1) + V(n)u(n) = z u(n)
\end{equation}
and is $\ell^2$ at $\pm \infty$.

Denote $\Z_+ = \{1,2,3,\ldots\}$ and $\Z_- = \{0,-1,-2,\ldots\}$. It is well known that
the maps $\mathcal{M}_\pm : V_\pm = V|_{\Z_\pm} \mapsto m_\pm$ are one-one and continuous
with respect to uniform convergence on compacta on the $m$-function side. The
$m$-functions $m_\pm$ are Herglotz functions, they have boundary values almost everywhere
on the real axis, and they are completely determined by their boundary values on any set
of positive Lebesgue measure.

We will be interested in those $V$ for which the functions $m_+,m_-$ obey identities like
\eqref{immpmeq} and \eqref{rempmeq}, that is,
\begin{equation}\label{mpmdet}
m_-(E+i0) = - \overline{m_+ (E+i0)}
\end{equation}
for a rich set of energies. Thus, for a set $\mathcal{Z} \subseteq \R$, we let
$$
\mathcal{D}(\mathcal{Z}) = \{ V \in R^\Z : m_\pm \text{ associated with $V$ obey
\eqref{mpmdet} for a.e. } E \in \mathcal{Z} \}.
$$
On $R^\Z$, define the shift transformation $[S(V)](n) = V(n+1)$.

\begin{lemma}\label{kottoplem}
Suppose that $\mathcal{Z} \subset \R$ has positive Lebesgue measure. Then:\\
{\rm (a)} $\mathcal{D}(\mathcal{Z})$ is $S$-invariant and closed in $R^\Z$.\\
{\rm (b)} For $V \in \mathcal{D}(\mathcal{Z})$, denote the restrictions to $\Z_\pm$ by
$V_\pm$. Then $V_-$ determines $V_+$ uniquely among elements of
$\mathcal{D}(\mathcal{Z})$
and vice versa.\\
{\rm (c)} If there exist $V^{(m)} , V \in \mathcal{D}(\mathcal{Z})$ such that $V^{(m)}_-
\to V_-$ pointwise, then $V^{(m)}_+ \to V_+$ pointwise.
\end{lemma}

\begin{proof}
(a) If $u_1,u_2$ denote the solutions of \eqref{evedetz} that obey $u_1(0) = u_2(1) = 1$
and $u_1(1) = u_2(0) = 0$, then we can write (note that we may normalize $u_\pm$ by
$u_\pm (0) = 1$)
$$
u_\pm(n) = u_1(n) \mp m_\pm (z) u_2(n).
$$
Let us denote the $m$-functions associated with $S(V)$ by $\tilde m_\pm$. Clearly,
$$
\tilde m_\pm(z) = \mp \frac{u_1(2) \mp m_\pm (z) u_2(2)}{u_1(1) \mp m_\pm (z) u_2(1)}.
$$
Since the $u_j(m)$ are polynomials in $z$ with real coefficients, this shows that
$$
\tilde m_-(E+i0) = - \overline{\tilde m_+ (E+i0)}
$$
for almost every $E \in \mathcal{Z}$ and hence $\mathcal{D}(\mathcal{Z})$ is
$S$-invariant. It follows from the continuity of the maps $\mathcal{M}_\pm$ that
$\mathcal{D}(\mathcal{Z})$ is closed. (For a proof that the identity between the boundary
values of the associated $m$-functions is preserved after taking limits, see \cite[Lemma~5]{k2}
and \cite[Lemma~7.4]{k3}.)\\
(b) $V_-$ determines $m_-$ and then \eqref{mpmdet} determines the boundary values of
$m_+$ on a set of positive Lebesgue measure. By general properties of Herglotz functions,
this determines $m_+$ (and hence $V_+$) completely. By the same argument, $V_+$ determines
$V_-$.\\
(c) By compactness, there is a subsequence of $\{V^{(m)}\}$ that converges pointwise,
that is, there is $\tilde V$ such that $V^{(m_k)} \to \tilde V$ as $k \to \infty$. By
part~(a), $\tilde V \in \mathcal{D}(\mathcal{Z})$. By assumption, $V_- = \tilde V_-$.
Thus, by part~(b), $V_+ = \tilde V_+$, and hence $V = \tilde V$. Consequently,
$V^{(m_k)}_+ \to V_+$ pointwise. In fact, we claim that $V^{(m)}_+ \to V_+$ pointwise.
Otherwise, we could reverse the argument (i.e., go from right to left) and show that
$V^{(\tilde m_k)}_- \not\to V_-$ for some other subsequence.
\end{proof}

We will now derive two important consequences of Lemma~\ref{kottoplem}: the absence of
absolutely continuous spectrum for topologically non-deterministic families and the
support theorem. To do so, we will consider the push-forward $\nu$ of $\mu$ on the
sequence space and its topological support.

More precisely, given an ergodic dynamical system $(\Omega,\mu,T)$ and a measurable
bounded sampling function $f : \Omega \to \R$ defining potentials $V_\omega(n) = f(T^n
\omega)$ as before, we associate the following dynamical system $(R^\Z,\nu,S)$: $R$ is a
compact set that contains the range of $f$, $\nu$ is the Borel measure on $R^\Z$ induced
by $\mu$ via $\Phi(\omega) = V_\omega$ (i.e., $\nu(A) = \mu(\Phi^{-1}(A))$), and $S$ is
the shift transformation on $R^\Z$ introduced above. Recall that the \textit{topological
support} of $\nu$, $\mathrm{supp} \, \nu$, is given by the intersection of all compact
sets $B \subseteq R^\Z$ with $\nu(B) = 1$. Clearly, $\mathrm{supp} \, \nu$ is closed and
$S$-invariant.

\begin{theorem}[Kotani 1989\footnote{The result appears explicitly in the 1989 paper \cite{k4}.
The main ingredients of the proof, however, were found earlier \cite{k2,
k3}.}]\label{kottopthm} Let $(\Omega,\mu,T,f)$ and $(R^\Z,d\nu,S)$ be as just described.
Assume that the set
$$
\mathcal{Z} = \{ E \in \R : \gamma(E) = 0 \}.
$$
has positive Lebesgue measure. Then,\\
{\rm (a)} Each $V \in \mathrm{supp} \, \nu$ is determined completely by $V_-$ {\rm
(}resp., $V_+${\rm )}.\\
{\rm (b)} If we let
$$
(\mathrm{supp} \, \nu)_\pm = \{ V_\pm : V \in \mathrm{supp} \, \nu \},
$$
then the mappings
\begin{equation}\label{extensionmaps}
E_\pm : (\mathrm{supp} \, \nu)_\pm \ni V_\pm \mapsto V_\mp \in (\mathrm{supp} \, \nu)_\mp
\end{equation}
are continuous with respect to pointwise convergence.
\end{theorem}

\begin{proof}
(a) By our earlier results, we know that $\mathcal{D}(\mathcal{Z})$ is compact and has
full $\nu$-measure. Thus, $\mathrm{supp} \, \nu \subseteq \mathcal{D}(\mathcal{Z})$ and
the assertion follows from Lemma~\ref{kottoplem}.(b).\\
(b) This follows from Lemma~\ref{kottoplem}.(c).
\end{proof}

We say that $(\Omega,\mu,T,f)$ is \textit{topologically deterministic} if there exist
continuous mappings $E_\pm : (\mathrm{supp} \, \nu)_\pm \to (\mathrm{supp} \, \nu)_\mp$
that are formal inverses of one another and obey $V^\#_- \in \mathrm{supp} \, \nu$ for
every $V_- \in (\mathrm{supp} \, \nu)_-$, where
$$
V^\#_-(n) = \begin{cases} V_-(n) & n \le 0, \\ E_-(V_-)(n) & n \ge 1. \end{cases}
$$
This also implies $V^\#_+ \in \mathrm{supp} \, \nu$ for every $V_+ \in (\mathrm{supp} \,
\nu)_+$, where
$$
V^\#_+(n) = \begin{cases} V_+(n) & n \ge 1, \\ E_+(V_+)(n) & n \le 0. \end{cases}
$$
Otherwise, $(\Omega,\mu,T,f)$ is \textit{topologically non-deterministic}.

\begin{coro}\label{topnd}
If $(\Omega,\mu,T,f)$ is \textit{topologically non-deterministic}, then
$\Sigma_\mathrm{ac} = \emptyset$.
\end{coro}

Our next application of Lemma~\ref{kottoplem} is the so-called support theorem; compare
\cite{k2}. For a Borel measure $\nu$ on $R^\Z$, let $\Sigma_\mathrm{ac}(\nu) \subseteq
\R$ denote the almost sure absolutely continuous spectrum, that is, $\sigma_\mathrm{ac}
(\Delta + V) = \Sigma_\mathrm{ac}(\nu)$ for $\nu$ almost every $V$. If $\nu$ comes from
$(\Omega,\mu,T,f)$, then $\Sigma_\mathrm{ac}(\nu)$ coincides with the set
$\Sigma_\mathrm{ac}$ introduced earlier. The support theorem says that
$\Sigma_\mathrm{ac}(\nu)$ is monotonically decreasing in the support of $\nu$.

\begin{theorem}[Kotani 1985]\label{suppthm}
For every $V \in \supp \, \nu$, we have $\sigma_\mathrm{ac} (\Delta + V) \supseteq
\Sigma_\mathrm{ac}(\nu)$. In particular, $\supp \, \nu_1 \subseteq \supp \, \nu_2$
implies that $\Sigma_\mathrm{ac} (\nu_1) \supseteq \Sigma_\mathrm{ac}(\nu_2)$.
\end{theorem}

\begin{proof}
We know that
$$
\supp \, \nu \subseteq \mathcal{D}(\mathcal{Z})  = \{ V \in R^\Z : m_\pm \text{
associated with $V$ obey \eqref{mpmdet} for a.e. } E \in \mathcal{Z} \}.
$$
Bearing in mind the Riccati equation \eqref{mminus}, a calculation like the one in
\eqref{hm} therefore shows that for every $V \in \supp \, \nu$, the Green function
associated with the operator $\Delta + V$ obeys $\Im G(0,0;E + i0) > 0$ for almost every
$E \in \mathcal{Z}$. This implies $\overline{\mathcal{Z}}^\mathrm{ess} \subseteq
\sigma_\mathrm{ac} (\Delta + V)$ and hence the result by Theorem~\ref{ipkthm}.
\end{proof}

A different proof may be found in Last-Simon \cite[Sect.~6]{ls}. Here is a typical
application of the support theorem:

\begin{coro}\label{ppgaps}
Let $\mathrm{Per}_\nu$ be the set of $V \in \supp \, \nu$ that are periodic, that is,
$S^p V = V$ for some $p \in \Z_+$. Then,
$$
\Sigma_\mathrm{ac}(\nu) \subseteq \bigcap_{V \in \mathrm{Per}_\nu} \!\!\sigma(\Delta +
V).
$$
\end{coro}

If there are sufficiently many gaps in the spectra of these periodic operators, one can
show in this way that $\Sigma_\mathrm{ac}(\nu)$ is empty.

Corollaries~\ref{topnd} and \ref{ppgaps} have been used in a variety of scenarios to
prove the absence of absolutely continuous spectrum. In fact, while the Kotani half of
Theorem~\ref{ipkthm} concerns the \textit{presence} of absolutely continuous spectrum on
$\mathcal{Z}$, it could be argued that the criteria for the \textit{absence} of
absolutely continuous spectrum that are by-products of the theory have been applied more
often. This is to a certain extent due to the fact that the majority of the ergodic
families of Schr\"odinger operators are expected to have no absolutely continuous
spectrum. The following ``conjecture'' is tempting because it is supported by a plethora
of results, both on the positive side and on the negative side. It has been verbally
suggested by Yoram Last and it has appeared explicitly in print in several places,
including \cite{j2,kk}.

\begin{problem}\label{apconjecture}
Show that $\mathrm{Leb} \, (\mathcal{Z}) > 0$ implies almost periodicity, that is, the
closure in $\ell^\infty(\Z)$ of the set of translates of $V_\omega$ is compact.
\end{problem}

Namely, the presence of (purely) absolutely continuous spectrum is known for all periodic
potentials, many limit-periodic potentials,\footnote{A sequence $V$ is limit-periodic if
there are periodic sequences $V^{(m)}$ such that $\|V - V^{(m)}\|_\infty \to 0$ as $m \to
\infty$. See, for example, Avron-Simon \cite{as3}, Chulaevsky \cite{chu1},
Chulaevsky-Molchanov \cite{cm}, and Pastur-Tkachenko \cite{pt}.} and some quasi-periodic
potentials (that are all uniformly almost periodic). On the other hand, the absence of
absolutely continuous spectrum is known for large classes of non-almost periodic ergodic
potentials. We will see some instances of the latter statement below. Nevertheless,
proving this conjecture is presumably very hard and it would already be interesting to
find further specific results that support the conjecture. For example, it is an open
(and seemingly hard) problem to prove the absence of absolutely continuous spectrum for
potentials defined by the skew shift.

We close this section with a problem concerning strips. As was mentioned at the end of
the Introduction, Kotani and Simon developed the analog of Kotani theory for discrete and
continuous Schr\"odinger operators with matrix-valued potentials in their 1988 paper
\cite{ks}. This framework includes in particular discrete Schr\"odinger operators on
strips. That is, operators of the form $\Delta + V_\omega$ on $\ell^2(\Z \times \{
1,\ldots,L \})$, where $\Delta$ is again given by the summation over nearest neighbors.
Transfer matrices are now $2L \times 2L$ and, modulo symmetry, there are $L$ Lyapunov
exponents, $\gamma_L(E) \ge \gamma_{L-1}(E) \ge \cdots \ge \gamma_1(E) \ge 0$. For the
general matrix-valued situation, they proved that the largest Lyapunov exponent is
positive for almost every energy if the potential is non-deterministic. This result
cannot be improved in this general setting. However, it is reasonable to expect that for
strips, the following result should hold.

\begin{problem}
Prove that for non-deterministic Schr\"odinger operators on a strip, all Lyapunov
exponents are non-zero for Lebesgue almost all energies.
\end{problem}

%
%
%
%

\section{Potentials Generated by the Doubling Map}

In this section we discuss potentials defined over the doubling map, that is, Example~(c)
from the introduction. The underlying dynamical system is strongly mixing and one would
hope that the spectral theory of the associated operators is akin to that of the Anderson
model, where the potentials are generated by independent, identically distributed random
variables. Alas, by dropping independence one loses the availability of most tools that
have proven useful in the study of the Anderson model.

While localization is expected for Schr\"odinger operators with potentials over the
doubling map, it has not been shown to hold in reasonable generality. There are only two
localization results in the literature, and each of them is to some extent
unsatisfactory. The first result was found by Bourgain and Schlag \cite{bs}, who proved
localization at small coupling and away from small intervals about the energies $\pm 2$
and $0$. Both assumptions seem unnatural. The other result is due to Damanik and Killip
\cite{dk2}, who proved localization for essentially all $f$ but only for Lebesgue almost
every boundary condition at the origin (recall that we are dealing with operators in
$\ell^2(\Z_+)$). A result holding for fixed boundary condition would of course be more
desirable. To this end, Damanik and Killip were at least able to show the absence of
absolutely continuous spectrum for fixed boundary condition in complete generality. These
results are indeed immediate consequences of Kotani theory and spectral averaging and we
give the short proofs below for the reader's convenience.

The first step in a localization proof for a one-dimensional Schr\"odinger operator is
typically a proof of positive Lyapunov exponents for many energies. For the Anderson
model, this can be done for all energies using F\"urstenberg's theorem or for Lebesgue
almost all energies using Kotani theory. At small coupling there is also a perturbative
approach due to Pastur and Figotin \cite{pf}. The extension of the approach based on
F\"urstenberg's theorem to potentials generated by the doubling map is not obvious; see,
however, \cite{ad3}. The perturbative approach extends quite nicely as shown by
Chulaevsky and Spencer \cite{cs}. Their results form the basis for the proof of the
partial localization result in \cite{bs}. Finally, the approach based on Kotani theory
also extends as we will now explain.

\begin{theorem}[Damanik-Killip 2005]\label{dkdmthm}
Suppose that $f \in L^\infty(\T)$ is non-constant and $V_\omega(n) = f(2^n \omega)$ for
$n \ge 1$. Then, the Lyapunov exponent $\gamma(E)$ is positive for Lebesgue almost every
$E \in \R$ and the absolutely continuous spectrum of the operator $H_\omega$ in
$\ell^2(\Z_+)$ is empty for Lebesgue almost every $\omega \in \T$.
\end{theorem}

\begin{proof}
Since the proof of this result is so short, we reproduce it here in its entirety. The
first step is to conjugate the doubling map $T$ to a symbolic shift via the binary
expansion. Let $\tilde \Omega_+ = \{0,1\}^{\Z_+}$ and define $D : \tilde \Omega_+ \to \T$
by $D(\omega) = \sum_{n = 1}^\infty \omega_n 2^{-n}$. The shift transformation, $S :
\tilde \Omega_+ \to \tilde \Omega_+$, is given by $(S \tilde \omega)_n = \tilde
\omega_{n+1}$. Clearly, $D \circ S = T \circ D$.

Next we introduce a family of whole-line operators as follows. Let $\tilde \Omega =
\{0,1\}^\Z$ and define, for $\tilde \omega \in \tilde \Omega$, the operator
$$
[H_{\tilde \omega} \phi](n) = \phi(n+1) + \phi(n-1) + V_{\tilde \omega}(n) \phi(n)
$$
in $\ell^2(\Z)$, where
$$
V_{\tilde \omega}(n) = f[D( \{ \tilde \omega_n, \tilde \omega_{n+1}, \tilde \omega_{n+2},
\ldots \} )].
$$
The family $\{H_{\tilde \omega}\}_{\tilde \omega \in \tilde \Omega}$ is non-deterministic
since $V_{\tilde \omega}$ restricted to $\Z_+$ only depends on $\{\tilde \omega_n\}_{n
\ge 1}$ and hence, by non-constancy of $f$, we cannot determine the values of $V_{\tilde
\omega}(n)$ for $n \le 0$ uniquely from the knowledge of $V_{\tilde \omega}(n)$ for $n
\ge 1$. It follows from Corollary~\ref{topnd} that the Lyapunov exponent for $\{H_{\tilde
\omega}\}_{\tilde \omega \in \tilde \Omega}$ is almost everywhere positive and
$\sigma_{{\rm ac}}(H_{\tilde \omega})$ is empty for almost every $\tilde \omega \in
\tilde \Omega$ with respect to the $(\frac{1}{2},\frac{1}{2})$-Bernoulli measure on
$\tilde \Omega$.

Finally, let us consider the restrictions of $H_{\tilde \omega}$ to $\ell^2(\Z_+)$, that
is, let $H_{\tilde \omega}^+ = E^* H_{\tilde \omega} E$, where $E : \ell^2(\Z_+) \to
\ell^2(\Z)$ is the natural embedding. Observe that $H_{\tilde \omega}^+ = H_\omega$,
where $\omega = D(\{\tilde \omega_1, \tilde \omega_2, \tilde \omega_2, \ldots\})$. This
immediately implies the statement on the positivity of the Lyapunov exponent for the
family $\{ H_\omega \}_{\omega \in \T}$. As finite-rank perturbations preserve absolutely
continuous spectrum, $\sigma_{{\rm ac}}(H_{\tilde \omega}^+) \subseteq \sigma_{{\rm
ac}}(H_{\tilde \omega})$ for every $\tilde \omega \in \tilde \Omega$. This proves that
$\sigma_{{\rm ac}}(H_{\tilde \omega}^+) = \emptyset$ for almost every $\tilde \omega \in
\tilde \Omega$.
\end{proof}

Given $\phi \in (-\pi/2,\pi/2)$, let $H_\omega^{(\phi)}$ denote the operator which acts
on $\ell^2(\Z_+)$ as in \eqref{oper}, but with $\psi(0)$ given by $\cos (\phi) \psi(0) +
\sin (\phi) \psi(1) = 0$. Thus, the original operator (with a Dirichlet boundary
condition) corresponds to $\phi = 0$. Theorem~\ref{dkdmthm} implies the following result
for this family of operators:

\begin{coro}
Suppose that $f$ is measurable, bounded, and non-constant. Then, for almost every $\phi
\in (-\pi/2,\pi/2)$ and almost every $\omega \in \T$, the operator $H_\omega^{(\phi)}$ in
$\ell^2(\Z_+)$ with potential $V_\omega(n) = f(2^n \omega)$ has pure point spectrum and
all eigenfunctions decay exponentially at infinity.
\end{coro}

\begin{proof}
This is standard and follows quickly from spectral averaging; see, for example,
\cite[Theorem~13.4]{pf} or \cite[Section~12.3]{simonti}.
\end{proof}

For a localization proof without the need for spectral averaging, it will be necessary to
prove the positivity of the Lyapunov exponent for a larger set of energies. Sufficient,
for example, is positivity away from a discrete set of exceptional energies. For
moderately small coupling, such a result will be contained in \cite{ad3}. The problem for
other values of the coupling constant is still open.

\begin{problem}
Find a class of functions $f \in L^\infty(\T)$ such that for every $\lambda \not= 0$, the
Lyapunov exponent associated with the potentials $V_\omega(n) = \lambda f(2^n \omega)$ is
positive away from a {\rm (}$\lambda$-dependent{\rm )} discrete set of energies.
\end{problem}

In connection with this problem, important obstructions have been found by Bochi \cite{b}
and Bochi and Viana \cite{bv}. Namely, positivity of $\gamma(E)$ away from a discrete set
will fail generically in $C(\T)$ (this result holds for rather general underlying
dynamics) and hence the H\"older continuity assumptions made in \cite{ad3} and \cite{cs}
are natural.

In some sense a large value of $\lambda$ alone should ensure the positivity of the
Lyapunov exponent. This is indeed the basis of several results for quasi-periodic
potentials or potentials generated by the skew-shift. For hyperbolic base transformations
such as the doubling map, however, there is a competition between two different kinds of
hyperbolic behavior that presents problems that have not been solved yet. Moreover, in
the large coupling regime, it would be especially interesting to prove uniform (in
energy) lower bounds on the Lyapunov exponents along with the natural $\log \lambda$
asymptotics.

\begin{problem}
Find a class of functions $f \in L^\infty(\T)$ such that for every $\lambda \ge
\lambda_0(g)$, the Lyapunov exponent associated with the potentials $V_\omega(n) =
\lambda f(2^n \omega)$ obeys $\inf_E \gamma(E) \ge c \log \lambda$ for some suitable
positive constant $c$.
\end{problem}

In this context it should be noted that Herman's subharmonicity proof for trigonometric
polynomials over ergodic shifts on the torus, \cite{h}, works in the case of the doubling
map.\footnote{The author is grateful to Kristian Bjerkl\"ov for pointing this out.} It
seems much less clear, however, how to carry over the Sorets-Spencer proof for
real-analytic $f$, \cite{ss}, from the case of irrational rotations of the circle to the
case of the doubling map, let alone the proof of Bourgain for real-analytic functions
over ergodic shifts on higher-dimensional tori \cite{b2}.

For $\mathrm{SL}(2,\R)$ cocycles over the doubling map that are not of Schr\"odinger form
(i.e., with a general $\mathrm{SL}(2,\R)$ matrix replacing \eqref{schcoc}), Young has
developed a method for proving positive Lyapunov exponents ``at large coupling'' that
works in the $C^1$ category \cite{y}. Her method does not immediately apply to
Schr\"odinger cocycles, but it would be interesting to find a suitable extension.

\begin{problem}
Modify Young's method and apply it to Schr\"odinger cocycles.
\end{problem}

%
%
%
%

\section{Absence of AC Spectrum for Rough Potentials}

It is in some way surprising that a rough sampling function $f$ can make the resulting
potentials non-deterministic. Traditionally, non-determinism had been thought of as a
feature induced by the underlying dynamics. In particular, quasi-periodic potentials (in
the generalized sense considered in this paper, which allows discontinuous $f$'s) had for
a long time been considered deterministic. The situation changed with an important
observation by Kotani in his short 1989 paper \cite{k4}.

He proved the following very general result:

\begin{theorem}[Kotani 1989]\label{kotthmfv}
Suppose that $(\Omega, T, \mu)$ is ergodic, $f : \Omega \to \R$ takes finitely many
values, and the resulting potentials $V_\omega$ are $\mu$-almost surely not periodic.
Then, $\mathrm{Leb} \, (\mathcal{Z}) = 0$ and therefore $\Sigma_\mathrm{ac} = \emptyset$.
\end{theorem}

In particular, operators with quasi-periodic potentials of the form
$$
V_\omega(n) = \lambda \sum_{m = 1}^N \gamma_m \chi_{[a_{m-1},a_m)}(n \alpha + \omega),
$$
where $0 = a_0 < a_1 < \cdots a_{N-1} < a_N = 1$, $\gamma_1,\ldots,\gamma_N \in \R$
(taking at leat two values) and $\lambda \not= 0$, have no absolutely continuous
spectrum. Note that this result holds for all non-zero couplings and hence it is
particularly surprising for small values of $\lambda$. We will have more to say about
these potentials in the next section.

The family $\{V_\omega\}_{\omega \in \T}$ does not seem to be non-deterministic in an
intuitive sense as $\omega$ is uniquely determined by the sequence $V_\omega |_{\Z_-}$.
However, the family becomes non-deterministic when we pass to the closed topological
support of the induced measure on $\{ \lambda \gamma_1,\ldots, \lambda \gamma_N \}^\Z$.
Then we will indeed find two distinct sequences that belong to the support of the induced
measure, whose restrictions to $\Z_-$ coincide. This fact will become more transparent
when we discuss Theorem~\ref{damkilthm1} below. Kotani's proof proceeded in a slightly
different way; he proved that there can be no \textit{continuous} mapping from the left
half-line to the right half-line.

\begin{proof}[Proof of Theorem~\ref{kotthmfv}.]
As above we denote the push-forward of $\mu$ under the map $\omega \mapsto V_\omega$ by
$\nu$. It suffices to show that $\mathrm{Leb} \, (\mathcal{Z}) > 0$ implies that $\supp
\nu$ is finite since then all elements of $\supp \nu$ are periodic.

By the continuity of the maps $E_\pm$ from \eqref{extensionmaps} and the fact that
$\mathrm{Ran} \, f$ is finite, there is a finite $M$ such that the knowledge of $V(-M) ,
\ldots, V(-1)$ determines $V(0)$ uniquely for $V \in \supp \nu$. Now shift and iterate!
It follows that $V(-M) , \ldots, V(-1)$ completely determine $\{ V(n) : n \ge 0 \}$ and
hence $\mathrm{supp} \, \nu$ has cardinality at most $(\# \, \mathrm{Ran} \, f)^M$.
\end{proof}

Consider the case where $\Omega$ is a compact metric space, $T$ is a homeomorphism, and
$\mu$ is an ergodic Borel probability measure. This covers most, if not all, applications
of interest. In this scenario, Damanik and Killip realized in \cite{dk} that finite range
of $f$ is not essential. What is important, however, is that $f$ is discontinuous at some
point $\omega_0 \in \Omega$. One can then use this point of discontinuity to actually
``construct'' two elements of $\supp \nu$ that coincide on a half-line.

We say that $l\in\R$ is an \textit{essential limit} of $f$ at $\omega_0$ if there exists
a sequence $\{\Omega_k\}$ of sets each of positive measure such that for any sequence
$\{\omega_k\}$ with $\omega_k \in \Omega_k$, both $\omega_k\to\omega_0$ and $f(\omega_k)
\to l$. If $f$ has more than one essential limit at $\omega_0$, we say that $f$ is
\textit{essentially discontinuous} at this point.

\begin{theorem}[Damanik-Killip 2005]\label{damkilthm1}
Suppose $\Omega$ is a compact metric space, $T : \Omega \to \Omega$ a homeomorphism, and
$\mu$ an ergodic Borel probability measure. If there is an $\omega_0 \in \Omega$ such
that $f$ is essentially discontinuous at $\omega_0$ but continuous at all points $T^n
\omega_0$, $n < 0$, then $\Sigma_\mathrm{ac} = \emptyset$.
\end{theorem}

\begin{proof}
We again denote the induced measure by $\nu$. For each essential limit $l$ of $f$ at
$\omega_0$, we will find $V_l \in \supp \nu$ with $V_l(0) = l$. By assumption and
construction, $V_l(n)$ is independent of $l$ for every $n < 0$. This shows $\mathrm{Leb}
\, (\mathcal{Z}) = 0$ and hence $\Sigma_\mathrm{ac} = \emptyset$.

Let $l$ be an essential limit of $f$ at $\omega_0$ and let $\{\Omega_k\}$ be a sequence
of sets which exhibits the fact that $l$ is an essential limit of $f$. Since each has
positive $\mu$-measure, we can find points $\omega_k \in \Omega_k$ so that $V_{\omega_k}$
is in $\supp \nu$; indeed, this is the case for almost every point in $\Omega_k$.

As $\omega_k \to \omega_0$ and $f$ is continuous at each of the points $T^n\omega_0$,
$n<0$, it follows that $V_{\omega_k}(n)\to V_{\omega_0}(n)$ for each $n<0$.  Moreover,
since $f(\omega_k)$ converges to $l$, we also have $V_{\omega_k}(0)\to l$. We can
guarantee convergence of $V_{\omega_k}(n)$ for $n>0$ by passing to a subsequence because
$R^\Z$ is compact.  Let us denote this limit potential by~$V_l$. As each $V_{\omega_k}$
lies in $\supp \nu$, so does $V_l$; moreover, $V(0)=l$ and $V_l(n) = V_{\omega_0}(n)$ for
each $n<0$.
\end{proof}

Here is an illustration of this result and a strengthening of the derived consequence:

\begin{coro}\label{damkilcoro1}
Suppose $\Omega = \T$, $\mu$ is normalized Lebesgue measure, and $T \omega = \omega +
\alpha$ for some irrational $\alpha$. If $f$ has a single {\rm (}non-removable{\rm )}
discontinuity at $\omega_0$, then for all $\omega\in[0,1)$, the operator $H_\omega$ has
no absolutely continuous spectrum.
\end{coro}

\begin{proof}
Let us say that $l$ is a limiting value of $f$ at $\omega_0$ if there is a sequence
$\{\omega_k\}$ in $\T \setminus \{\omega_0\}$ such that $\omega_k \to \omega_0$ and
$f(\omega_k)\to l$. As $f$ has a non-removable discontinuity at $\omega_0$, it has more
than one limiting value at this point. Moreover, since $f$ is continuous away from
$\omega_0$, any limiting value is also an essential limit since we can choose each
$\Omega_k$ to be a suitably small interval around $\omega_k$.

This shows that $f$ has an essential discontinuity at $\omega_0$. As the orbit of
$\omega_0$ never returns to this point, $f$ is continuous at each point $T^n \omega_0$,
$n \neq 0$. Therefore, Theorem~\ref{damkilthm1} is applicable and shows that $H_\omega$
has no absolutely continuous spectrum for Lebesgue-almost every $\omega\in[0,1)$.

It remains to show that the absolutely continuous spectrum of $H_\omega$ is empty for
\textit{all} $\omega \in \R / \Z$. We begin by fixing $\omega_1$ such that $H_{\omega_1}$
has no absolutely continuous spectrum and such that the orbit of $\omega_1$ does not meet
$\omega_0$; almost all $\omega_1$ have these properties.

Given an arbitrary $\omega\in\R/\Z$, we may choose a sequence of integers $\{n_i\}$ so
that $T^{n_i}(\omega) \to \omega_1$.  As $f$ is continuous on the orbit of $\omega_1$,
the potentials associated to $T^{n_i}(\omega)$ converge pointwise to $V_{\omega_1}$. By a
result of Last and Simon \cite{ls}, the absolutely continuous spectrum cannot shrink
under pointwise approximation using translates of a single potential. Thus, the operator
with potential $V_\omega$ cannot have absolutely continuous spectrum. This concludes the
proof.
\end{proof}

These results are particularly interesting in connection with Problem~\ref{apconjecture}.
Quasi-periodic potentials (as defined in this paper) are almost periodic if and only if
$f$ is continuous. Thus, proving the absence of absolutely continuous spectrum for
quasi-periodic potentials with discontinuous $f$'s is a way of providing further support
for the conjecture that $\Sigma_\mathrm{ac} \not= \emptyset$ implies almost periodicity.
\\[1mm]

Let us now turn to the case of continuous sampling functions $f$. The proof of
Theorem~\ref{damkilthm1} certainly breaks down and it is not clear where some sort of
non-determinism should come from in the quasi-periodic case, for example. Of course,
absence of absolutely continuous spectrum does not hold for a general continuous $f$.
Thus, the following result from \cite{ad} is somewhat surprising:

\begin{theorem}[Avila-Damanik 2005]\label{avdamthm1}
Suppose $\Omega$ is a compact metric space, $T : \Omega \to \Omega$ a homeomorphism, and
$\mu$ a non-atomic ergodic Borel probability measure. Then, there is a residual set of
functions $f$ in $C(\Omega)$ such that $\Sigma_{{\rm ac}}(f) = \emptyset$.
\end{theorem}

Recall that a subset of $C(\Omega)$ is called residual if it contains a countable
intersection of dense open sets. A residual set is locally uncountable.

One would expect some absolutely continuous spectrum for weak perturbations with
sufficiently nice potentials; especially in the one-frequency quasi-periodic case. More
precisely, if $f$ is nice enough, then $\Delta + \lambda f(n\alpha + \omega)$ should have
some/purely absolutely continuous spectrum for $|\lambda|$ sufficiently small. It is
known that real-analyticity is sufficiently ``nice enough'' \cite{bj} (when $\alpha$ is
Diophantine), but it was expected that this assumption is much too strong and could
possibly be replaced by mere continuity. The proof of Theorem~\ref{avdamthm1} can easily
be adapted to yield the following result, also contained in \cite{ad}, which shows that
continuity of the sampling function is not sufficient to ensure the existence of
absolutely continuous spectrum for weakly coupled quasi-periodic potentials.

\begin{theorem}[Avila-Damanik 2005]\label{avdamthm2}
Suppose $\Omega$ is a compact metric space, $T : \Omega \to \Omega$ a homeomorphism, and
$\mu$ a non-atomic ergodic Borel probability measure. Then, there is a residual set of
functions $f$ in $C(\Omega)$ such that $\Sigma_{{\rm ac}}(\lambda f) = \emptyset$ for
almost every $\lambda > 0$.
\end{theorem}

\begin{proof}
We only sketch the proofs of Theorems~\ref{avdamthm1} and \ref{avdamthm2}. The key
technical issue is to establish that the maps
\begin{equation} \label {L11}
(L^1(\Omega) \cap B_r(L^\infty(\Omega)), \| \cdot \|_1) \to \R, \; \; f \mapsto
\mathcal{Z}(f)
\end{equation}
and
\begin{equation} \label {L12}
(L^1(\Omega) \cap B_r(L^\infty(\Omega)), \| \cdot \|_1) \to \R, \; \; f \mapsto
\int_0^\Lambda \mathcal{Z}(\lambda f) \, d\lambda
\end{equation}
are upper semi-continuous. Here, $\Lambda > 0$, $B_r(L^\infty(\Omega) = \{ f \in
L^\infty(\Omega) : \|f\|_\infty < r \}$, and $\mathcal{Z}(f)$ denotes the set of energies
for which the Lyapunov exponent associated with $(\Omega,T,\mu,f)$ vanishes.

Upper semi-continuity of the map \eqref{L11} can be shown using the fact that $\gamma$ is
harmonic in the upper half-plane and subharmonic on the real line; see \cite{ad} for
details. Fatou's Lemma then implies upper semi-continuity of \eqref{L12}.

For $\delta > 0$, define
$$
M_\delta = \{ f \in C(\Omega) : \mathrm{Leb}(\mathcal{Z}(f)) < \delta \}.
$$
By the upper semi-continuity statement above, $M_\delta$ is open. By approximation with
discontinuous functions and upper semi-continuity again, we see that $M_\delta$ is also
dense.

It follows that
\begin{align*}
\{ f \in C(\Omega) : \Sigma_{{\rm ac}}(f) = \emptyset \} & = \{ f \in C(\Omega) :
\mathrm{Leb}(\mathcal{Z}(f)) = 0 \} = \bigcap_{\delta > 0} M_\delta
\end{align*}
is residual and Theorem~\ref{avdamthm1} follows. Given upper semi-continuity of
\eqref{L12}, the proof of Theorem~\ref{avdamthm2} is analogous.
\end{proof}

While continuous functions can be approximated in the $C^0$ norm by discontinuous
functions, this does not work in the $C^\varepsilon$ norm for any $\varepsilon > 0$.
Thus, the proof just given does not extend to H\"older classes. It would be interesting
to explore possible extensions of the results themselves; thus motivating the following
problem.

\begin{problem}
Prove or disprove statements like the ones in Theorems~\ref{avdamthm1} and
\ref{avdamthm2} for H\"older classes $C^\varepsilon(\Omega)$, $\varepsilon > 0$.
\end{problem}

%
%
%
%

\section{Uniform Lyapunov Exponents and Zero-Measure Spectrum}\label{secfv}

The Kotani result for potentials taking finitely many values, Theorem~\ref{kotthmfv}, is
central to the study of one-dimensional quasi-crystal models. The main results in this
area have been reviewed in \cite{d1,d2,s7}. In this section we will therefore focus on
the recent progress and discuss why zero-measure spectrum is a consequence of Kotani
theory when there is uniform convergence to the Lyapunov exponent.

One-dimensional quasi-crystals are typically modelled by sequences over a finite alphabet
which are aperiodic but which have very strong long-range order properties. An important
class of examples is given by one-frequency quasi-periodic potentials with step functions
as sampling functions. That is, the potentials are of the form
\begin{equation}\label{fvqppot}
V_\omega(n) = \lambda \sum_{m = 1}^N \gamma_m \chi_{[a_{m-1},a_m)}(n \alpha + \omega),
\end{equation}
where $0 = a_0 < a_1 < \cdots < a_N = 1$ is a partition of the unit circle, $\lambda,
\gamma_1, \ldots, \gamma_N$ are real numbers, $\alpha$ is irrational, and $\omega \in
\T$. We obtain aperiodic potentials if $\lambda \not= 0$ and $\{\gamma_1, \ldots,
\gamma_N\}$ has cardinality at least two. We will assume these conditions throughout this
section.

One of the properties that has been established for many quasi-crystal models is
zero-measure spectrum. By general principles, this implies that the spectrum is a Cantor
set because it cannot contain isolated points. Given the Kotani result, $\mathrm{Leb} \,
(\mathcal{Z}) = 0$, the natural way of proving this is via the identity $\Sigma =
\mathcal{Z}$.

\begin{theorem}[Damanik-Lenz 2006]\label{damlenthm}
Suppose the potentials are of the form \eqref{fvqppot} and in addition all discontinuity
points $\{ a_m \} \subset \T$ are rational. Then, the Lyapunov exponent vanishes
identically on the spectrum, that is, $\Sigma = \mathcal{Z}$. As a consequence, the
spectrum is a Cantor set of zero Lebesgue measure.
\end{theorem}

\begin{proof}
We only sketch the main ideas. More details can be found in \cite{dl,dl2}. Denote
$$
\mathcal{UH} = \left\{ E : \tfrac{1}{n} \log \| A^E_n (\omega) \| \to \gamma(E) > 0
\text{ uniformly in } \omega \right\}.
$$
Then, by Lenz \cite{len2} (see also Johnson \cite{john}), $\mathcal{UH} = \C \setminus
\Sigma$. In particular,
$$
\Sigma = \mathcal{Z} \cup \mathcal{NUH},
$$
where
$$
\mathcal{NUH} = \left\{ E : \gamma(E) > 0 \text{ and } E \not\in \mathcal{UH} \right\}.
$$
The result follows once $\mathcal{NUH} = \emptyset$ is established. Thus, given $E$ with
$\gamma(E) > 0$, we need to show that $\tfrac{1}{n} \log \| A^E_n (\omega) \| \to
\gamma(E)$ uniformly in $\omega$.

Uniform convergence along a special subsequence, $n_k \to \infty$, can be shown using
results from Boshernitzan \cite{bosh} and Lenz \cite{len1}. Namely, the assumption that
all $a_m$ are rational implies that there is a sequence of integers $n_k \to \infty$ such
that for each $k$, all words of length $n_k$ that occur in the potentials $V_\omega$ do
so with comparable frequencies That is, there is a uniform $C > 0$ such that for every
$k$,
\begin{equation}\label{bcond}
\min_{|w| = n_k, w \text{ occurs}} \liminf_{J \to \infty} \frac1J \# \{ j : 1 \le j \le
J, \; V_\omega(j) \ldots V_\omega(j + n_k - 1) = w \} \ge \frac{C}{n_k}
\end{equation}
uniformly in $\omega$ \cite{bosh}. Using this result, one can then use ideas from
\cite{len1} to show that $\tfrac{1}{n_k} \log \| A^E_{n_k} (\omega) \| \to \gamma(E)$ as
$k \to \infty$, uniformly in $\omega$.

Finally, the avalanche principle of Goldstein and Schlag \cite{gs} allows one to
interpolate and prove the desired uniform convergence of $\tfrac{1}{n} \log \| A^E_n
(\omega) \|$ to $\gamma(E)$ as $n \to \infty$.
\end{proof}

The rationality assumption in Theorem~\ref{damlenthm} holds on a dense set of parameters,
which makes it suitable for an approximation of a continuous sampling function by a
sequence of step functions. Some consequences that may be drawn from this can be found in
Bjerkl\"ov et al.\ \cite{bdj}. On the other hand, the assumption is certainly not
necessary and more general results than the one presented here can be found in
\cite{dl2}. It would be nice if the assumption could be removed altogether:

\begin{problem}
Prove zero-measure spectrum for all finite partitions of the circle, that is, remove the
rationality assumption from Theorem~\ref{damlenthm}.
\end{problem}

It should be mentioned, however, that the proof sketched above will not work in this
generality. The approach is based on the Boshernitzan condition \eqref{bcond}, and it was
shown in \cite{dl2} that this condition fails for certain parameter values.

From a mathematical point of view, the following problem is natural:

\begin{problem}
Study multi-frequency analogs. That is, for a finite partition $\T^d = J_1 \cup \cdots
\cup J_N$ and the operators with potential
$$
V_\omega(n) = \lambda \sum_{m = 1}^N \gamma_m \chi_{J_m}(n \alpha + \omega),
$$
what is the measure of the spectrum and what is the spectral type?
\end{problem}

These potentials are not directly motivated by quasi-crystal theory but they form an
interesting class that may again prove useful in the understanding of the phenomena that
arise for continuous sampling functions. Moreover, very little is understood about the
associated operators apart from the general Kotani result, which says that there is never
any absolutely continuous spectrum. It is unclear, however, whether there can be any
point spectrum, for example.

%
%
%
%

\section{Purely AC Spectrum for the Subcritical AMO}

In this final section, we describe the (to the best of our knowledge) first application
of Corollary~\ref{pureaccoro}. In his 1997 paper, Kotani writes that at the time Kotani
theory was developed, ``it was not clear whether we could know the pure absolute
continuity ...\ only from the IDS, but this corollary has answered this question
affirmatively.''

The application we will present involves the almost Mathieu operator
$$
[H_\omega \psi](n) = \psi(n+1) + \psi(n-1) + 2 \lambda \cos (2\pi (n\alpha + \omega))
\psi(n),
$$
that is, the Schr\"odinger operator with one-frequency quasi-periodic potential
associated with the sampling function $f(\omega) = 2\lambda \cos (2\pi \omega)$. This
operator is known to exhibit a metal-insulator transition at $|\lambda| = 1$, that is,
for almost every $\alpha$, the almost sure spectral type is purely absolutely continuous
for $|\lambda| < 1$, purely singular continuous for $|\lambda| = 1$, and pure point (with
exponentially decaying eigenfunctions) for $|\lambda| > 1$. See Jitomirskaya \cite{j} for
this result and its history. For $|\lambda| > 1$, one indeed has to exclude a
zero-measure set of frequencies $\alpha$ since it was shown by Avron and Simon, using
Gordon's Lemma, that for Liouville $\alpha$, there are no eigenvalues \cite{as2,as}. For
$|\lambda| \le 1$, it has long been expected that the results above extend to all
irrational frequencies. For $|\lambda| = 1$, the issue was resolved by Avila and
Krikorian \cite{avilak}. The question of what happens for $|\lambda| < 1$ was addressed
by Problem~6 in \cite{s21st}.

Since Bourgain and Jitomirskaya showed in \cite{bj2} that the Lyapunov exponent
associated with the almost Mathieu operator obeys $\gamma(E) = \max \{ 0, \log | \lambda
| \}$ for every $E \in \Sigma$ for all irrational frequencies $\alpha$, the problem
reduces to a study of the integrated density of states, that is, to a proof of its
absolute continuity. The following result was shown in \cite{ad2} and it completely
settles this regularity issue for the integrated density of states.

\begin{theorem}[Avila-Damanik 2006]\label{avdamthm3}
The integrated density of states of the almost Mathieu operator is absolutely continuous
if and only if $|\lambda| \not = 1$.
\end{theorem}

Combining this result with the one from \cite{bj2} just quoted along with
Corollary~\ref{pureaccoro}, we obtain almost surely purely absolutely continuous spectrum
for the subcritical (i.e., $|\lambda| < 1$) almost Mathieu operator:

\begin{coro}
If $|\lambda| < 1$, then $\Sigma_\mathrm{sing} = \emptyset$.
\end{coro}

Subsequently, Avila even extended this result and proved purely absolutely continuous
spectrum for $|\lambda| < 1$, $\alpha$ irrational, and \textit{every} $\omega$ \cite{a}.

\begin{problem}
Extend the results of this section to more general $f \in L^\infty(\T)$; for example,
real-analytic $f$'s.
\end{problem}

It was shown by Bourgain and Jitomirskaya that for Diophantine frequency $\alpha$ and
analytic $f$, $\Sigma_\mathrm{sing}(\lambda f) = \emptyset$ for $\lambda$ sufficiently
small \cite{bj}. An extension of this result to all $\omega$ is contained in \cite{aj}.
The proofs probably break down for Liouville frequencies (for this, one needs to extend
the method based on Gordon's Lemma to matrices that are not banded, but which do have
exponential off-diagonal decay). Thus, it seems natural to attack the problem for
Liouville frequencies in the same way as above.

\end{document}